\documentclass[useAMS,usenatbib,times,letter,amssymb]{mn2e}
\usepackage{epsfig,times,amssymb,amsmath,xcolor,verbatim}
\usepackage{xspace}

\usepackage{natbib,hyperref,ifthen} 
\def\eprinttmp@#1arXiv:#2 [#3]#4@{ 
\ifthenelse{\equal{#3}{x}}{\href{http://arxiv.org/abs/#1}{#1} 
}{\href{http://arxiv.org/abs/#2}{arXiv:#2} [#3]}} 

\newcommand{\eprint}[1]{\eprinttmp@#1arXiv: [x]@}

\title[Simultaneous Constraints on Cosmology and Photometric Redshifts from Weak Lensing and Galaxy Clustering]{Simultaneous Constraints on Cosmology and Photometric Redshift Bias from Weak Lensing and Galaxy Clustering}

\author[S. Samuroff et al]{S. Samuroff$^{1}$\thanks{\texttt{simon.samuroff@postgrad.manchester.ac.uk}}, M.A. Troxel$^1$, S.L. Bridle$^1$, J. Zuntz$^1$, N. MacCrann$^1$, E. Krause$^{2}$, 
\newauthor{
T. Eifler$^{3,4}$, D. Kirk$^5$}
\\
$^1$Jodrell Bank Centre for Astrophysics, University of Manchester, Oxford Road, Manchester, M13 9PL, UK. \\ 
$^2$Kavli Institute for Particle Cosmology and Astrophysics, Stanford University, Stanford, CA 94305, USA\\
$^3$Jet Propulsion Laboratory, California Institute of Technology, Pasadena, CA 91109, USA\\
$^4$Department of Physics, California Institute of Technology, Pasadena, CA 91125, USA\\
$^5$Department of Physics and Astronomy, University College London, Gower Street, London WC1E 6BT, UK.\\
}

\newcommand{\be}{\begin{equation}}  \newcommand{\ee}{\end{equation}}
  
\newcommand{\ba}{\begin{eqnarray}}\newcommand{\ea}{\end{eqnarray}}

\newcommand{\pz}{photo-$z$\xspace}

\newcommand{\blockfont}[1]{{\textsc{#1}}\xspace}

\def\gs{\mathrel{\raise1.16pt\hbox{$>$}\kern-7.0pt %
\lower3.06pt\hbox{{$\scriptstyle \sim$}}}}         %
\def\ls{\mathrel{\raise1.16pt\hbox{$<$}\kern-7.0pt %
\lower3.06pt\hbox{{$\scriptstyle \sim$}}}}         %

\begin{document}

\maketitle

\begin{abstract}
We investigate the expected cosmological constraints from a combination of weak lensing and large-scale galaxy clustering using realistic redshift distributions. 
Introducing a systematic bias in the weak lensing redshift distributions (of 0.05 in redshift) produces a $>2\sigma$ bias in the recovered matter power spectrum amplitude and dark energy equation of state, for preliminary Stage III surveys.
We demonstrate that these cosmological errors can be largely removed by marginalising over unknown biases in the assumed weak lensing redshift distributions, if we assume high quality redshift information for the galaxy clustering sample.
Furthermore the cosmological constraining power is mostly retained despite removing much of the information on the weak lensing redshift distribution biases.
We show that this comes from complementary degeneracy directions between cosmic shear and the combination of galaxy clustering with cross-correlation between shear and galaxy number density.
Finally we examine how the self-calibration performs when the assumed distributions differ from the true distributions by more than a simple uniform bias. We find that the effectiveness of this self-calibration method will depend on the details of a given experiment and the nature of the uncertainties on the estimated redshift distributions.\end{abstract}



\section{Introduction}

Cosmic shear is potentially the most powerful tool available to cosmologists today. As an unbiased probe of the mass distribution, it offers powerful constraints on the mean density of the Universe and the clustering of dark matter.
It is also expected to shed new light on the late-time accelerated expansion of the Universe and thus measure the dark energy equation of state and test General Relativity on the largest scales. 

\begin{figure}
   \centering
   \includegraphics[width=9cm]{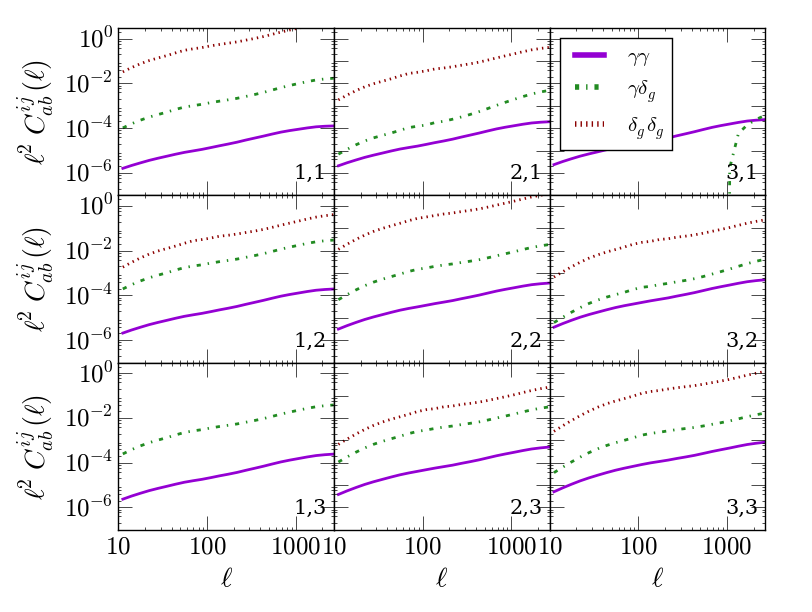}
   \caption{The components of the fiducial datavector used in this Letter. Shown are angular power spectra of cosmic shear (purple solid), galaxy clustering (red dotted) and the shear-density cross-correlation (green dot dashed). Each panel corresponds to a unique redshift bin pairing.  In the panels where it is not visible, the $\delta _g \delta _g$ spectrum is below the range shown. We note that all values shown are positive, apart from $C^{1,3}_{\gamma \delta _g}$ (upper right), which becomes negative and is below the lowest point on this scale for $\ell<900 $.}
   \label{fig:cls}
\end{figure}

A three decade programme aiming to extract unprecedented constraints on our cosmological model from 
cosmic shear is now midway to completion. It began soon after the first detection in 2000 \citep{bacon00,vanWaerbeke00,wittman00,kaiser_detection} using $\sim$10000 galaxies and is expected to culminate in catalogues of more than a billion galaxies by the end of the coming decade (Stage IV, \citet{albrecht06}). Logarithmically, we are halfway there, with ongoing analyses of the preliminary Stage III datasets, containing $\sim$10 million galaxies (\citealt{dessv2pt15,kids16}, see also \citealt{heymans13,jee16}).
The increase in the number of galaxies with reliable shape measurements has allowed tighter cosmology constraints, but also requires better control of systematic biases.
In this Letter we focus on a potential Achilles' heel of galaxy imaging surveys for cosmology: the use of photometric redshifts to estimate distances to galaxies. 

Tomographic cosmic shear analyses bring a number of benefits \citep{hu99}, but place stringent requirements on our knowledge of galaxy redshift distributions. 
\citet{amara07,abdalla08,jouvel09} and \citet{ishak06} present detailed studies of the requirements for  spectroscopic follow up of
Stage IV cosmology surveys, while \citet{ma06,huterer06, bernstein09} offer numerical forecasts of cosmological impact from
photometric redshift (\pz) biases. Many others (e.g. \citealt{bordoloi12,cunha_huterer12}) present detailed studies of specific \pz systematics, albeit with less focus on the ultimate cosmological impact.
Tightening systematics requirements have sparked interest in spatial cross-correlations between photometric and spectroscopic galaxies within the survey volume, as a method for externally calibrating photometric redshifts \citep{newman08,menard13,dePutter14,rahman15,choi15,scottez16}. 

\begin{figure}
   \centering
   \includegraphics[width=1\columnwidth]{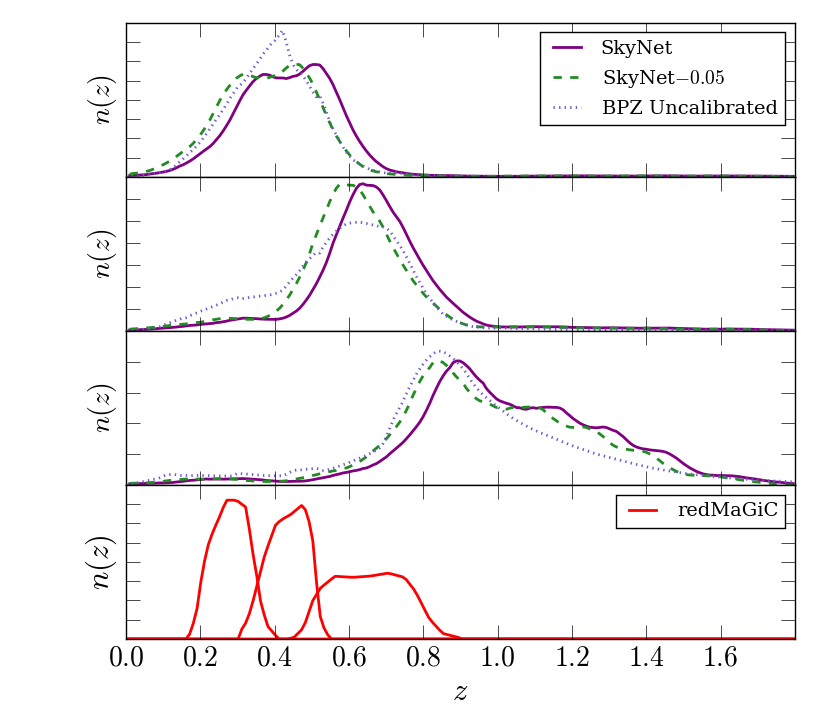}
   \caption{
Redshift distributions considered in this paper. The upper panels show the shear catalogue redshift distributions used in this work, taken from DES SV \citep{bonnettsv15}:
\blockfont{skynet} (solid purple; fiducial), \blockfont{skynet} with a bias of 0.05 (dashed green) and \blockfont{bpz} (dotted blue) without the shift of 0.05 in redshift used in \citet{bonnettsv15}.
The lower panel displays the the fiducial galaxy clustering catalogue in three bins
taken from~\citet{clampitt16} which uses~\citet{rozo15} (DES SV redMaGiC).
}   
\label{fig:nofzs}
\end{figure}

\begin{figure*}
   \centering
   (a)
\includegraphics[width=1\columnwidth]{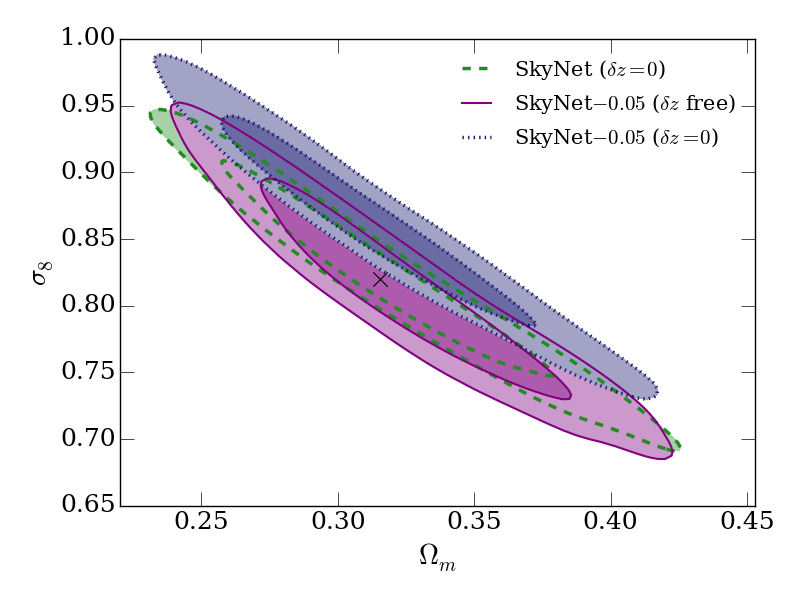}
(b)
\includegraphics[width=1\columnwidth]{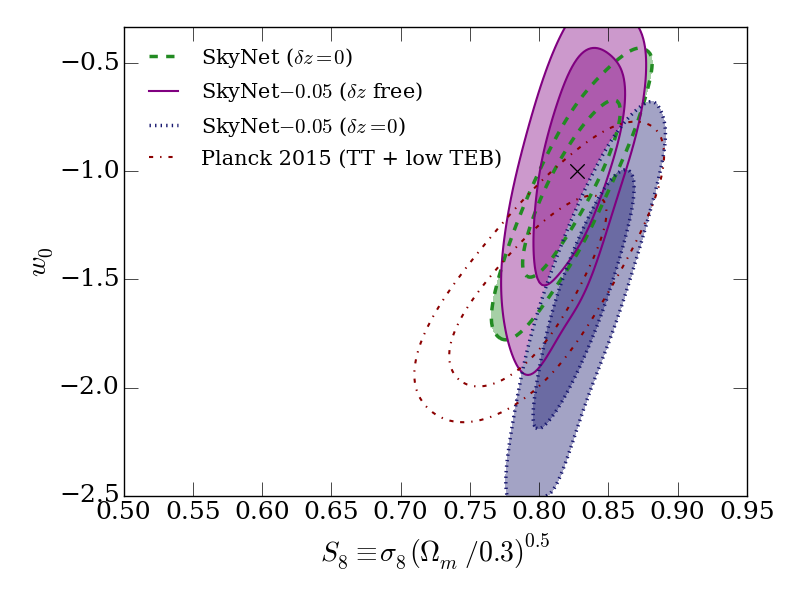}
   \caption{
(a) Forecast constraints on the matter density and clustering amplitude in $\Lambda$CDM and (b) dark energy equation of state in $w$CDM for various assumptions about photometric redshifts (for reference constraints from Planck 2015 temperature and low frequency polarisation data alone are shown by the dot-dashed red contours \citep{planck2015cosmo}). 
The colours in each panel indicate three \pz scenarios. In green are the result of using the \blockfont{skynet} $n(z)$ in the theory calculation data and fixing $\delta z=0$. We show this as an unrepresentative ideal case, where the photometric estimates provide a perfect representation of the true galaxy distribution. Overlain are the results of using the \blockfont{skynet} $n(z)$, biased downwards by 0.05 in redshift
under the (erroneous) assumption of no bias (blue dotted) and varying three additional $\delta z^i$ nuisance parameters marginalised with a wide Gaussian prior of width $\Delta \delta z =0.1$(purple solid). 
The true input cosmology is shown by the black cross.
}   
\label{fig:biases}
\end{figure*}

Given the limited amount of spectroscopic information available, several authors have speculated about the possibility of calibrating cosmic shear redshift distributions from the imaging survey itself. \citet{huterer06} show that cosmic shear alone affords a limited capacity for self-calibration.
\citet{schneider06} and \citet{sun15} investigate the photometric redshift calibration information available from Stage IV galaxy clustering.
\citet{zhan06} explore the constraining power on $w_0$ using a similar technique 
and combine with cosmic shear constraints. \citet{zhang10} point out that shear-density cross-correlations (cross-correlation between shear and galaxy counts, also referred to as tangential shear or galaxy-galaxy lensing) can help to constrain \pz error, when combined with galaxy clustering. 

All the studies mentioned in the previous paragraph make a crucial assumption, which is unlikely to be realised in practice: that the galaxies used for cosmic shear have a systematics-correctable galaxy clustering signal. In practice regions of the sky with better (worse) seeing conditions are likely to contain a higher (lower) number density of galaxies usable for cosmic shear \citep[e.g. see Appendix C of][for more discussion]{choi15}. Therefore there will be a large spurious clustering signal from the galaxies selected for a shear catalogue, rendering standard galaxy clustering analyses useless. Thus in practice we will usually have a different galaxy sample selection for the weak lensing sample and the galaxy clustering sample. 
This is standard in galaxy-galaxy lensing analyses and was done in the first combined analyses of cosmic shear and large scale structure on data (\citealt{nicola16}, who also combined with the CMB), and was considered for Stage IV surveys with much tighter priors in the forecasts of \citet{cosmolike16}. 
This means one has twice as many redshift distributions to understand as in a shear-only analysis. However, this also offers an opportunity: we can choose to use a galaxy clustering sample with much better understood photometric redshift properties, which can in turn help to calibrate the redshift distribution of the weak lensing sample. 

In this Letter we explore the potential for simultaneously constraining \pz error and cosmology using cosmic shear, galaxy clustering and shear-density cross-correlations. Unlike previous studies using this data combination, we consider a scenario in which the redshift distribution of the shear catalogue and galaxy clustering catalogues differ significantly. We assume the galaxy clustering 
sample is highly homogeneous and dominated by luminous red galaxies, which tend to yield high quality \pz. 

This Letter is structured as follows.
Section 2 outlines the setup of our analysis with a description of the simulated data vectors, redshift distributions and the photometric uncertainties considered. 
In Section 3 we investigate the power of 
cosmic shear, galaxy clustering and shear-density cross-correlations
to internally constrain \pz biases.
Finally a series of robustness tests are presented to explore the limits of this effect. We adopt a 
fiducial flat $\Lambda$CDM cosmology with $\sigma_8=0.82$, $\Omega_{\rm m} = 0.32$, $h=0.67$, $w_0 =-1$, $\Omega_{\rm b} = 0.049$.


\section{Methodology and assumptions}

We follow a method similar to \citet{jb10} to implement a forecast of the three weak lensing plus large-scale structure two-point functions: cosmic shear, galaxy clustering and shear-density cross-correlations. 
We carry out a regular MCMC forecast by simulating a datavector and covariance from the fiducial cosmology and then generating a list of acceptable cosmologies by fitting trial cosmologies to the simulated datavector.
The fiducial datavector, illustrated in Fig. 1, contains three types of angular power spectrum, each with 25 logarithimically spaced top hat bins over the range $10 < \ell < 3000$. 
We use \blockfont{CosmoSIS}
\citep{zuntz13}
to MCMC sample parameter space and compute matter power spectra using 
\blockfont{camb} \citep{camb} with nonlinear corrections from \citet{takahashi12}. 


The fiducial analysis assumes a galaxy catalogue typical of the size of a preliminary Stage III survey.
To be specific, we use the Dark Energy Survey Science Verification (DES SV) galaxy number density given in \citet{jarvis15} of 6.8 galaxies per square arcminute, with $\sigma_{\epsilon}=0.2$. and an area of 1500 square degrees gives 37M galaxies in total, which is a little larger than, or comparable to, CFHTLenS\footnote{http://www.cfhtlens.org}, and to KiDS\footnote{http://kids.strw.leidenuniv.nl}, DES\footnote{http://www.darkenergysurvey.org}and HSC\footnote{http://www.naoj.org/Projects/HSC} main survey preliminary analyses.
We use the \blockfont{skynet} $n(z)$ used in the DES SV weak lensing analyses presented in \citet{bonnettsv15}.
We marginalise over multiplicative shear calibration uncertainty with a Gaussian prior ($\Delta m=0.02$).
\citep[see also][]{jarvis15, fenechconti16, jee16}.
To be conservative in our cosmology constraints, we model intrinsic alignments with the commonly used nonlinear alignment model \citep{bk07} with an additional power law in redshift (e.g. \citealt{joachimi11,dessv2pt15}) and allow the GI and II amplitude and power law to be different, giving four free parameters in total.

To model a realistic galaxy clustering catalogue the $n(z)$ of the DES 
SVredMaGiC luminous red galaxy
catalogue \citep{rozo15} is adopted from \citet{clampitt16}.
A linear galaxy bias is applied in each bin and marginalised over with a wide flat prior.
To avoid the non-linear galaxy bias regime we impose conservative scale cuts to the clustering sample. 
The minimum scale used in each tomographic bin is determined by rescaling the prescription presented in \citet{rassat08} to match, where available, the minimum scales determined by \citet{kwan16} to be unaffected by nonlinear biasing, which leads to multiplying the \citet{rassat08} cuts by a factor of three.

To parameterise uncertainties in the redshift distributions a bias $\delta z^i$ is applied, describing a uniform linear translation, $\tilde{n}^{i}(z) = n^{i}( z+\delta z^{i} )$.
For the galaxy clustering sample we marginalise over $\delta z^i$ with a Gaussian prior of standard deviation $\Delta \delta z = 0.01$. 
The \pz for the shear catalogue are somewhat lower in quality due to the number and type of objects required for shear measurement and we apply a conservative Gaussian prior ($\Delta \delta z = 0.1$) for the fiducial analysis. 

In addition to the nuisance parameters described above, we leave 5 cosmological parameters free to vary with no external priors. The fiducial analysis then has 21 degrees of freedom, $\textbf{p}=(\sigma_8, \Omega_{\rm m}, h, \Omega_{\rm b}, n_s, A_{GI}, A_{II}, \eta_{GI}, \eta_{II}, m^i, \delta z^i, b^j_g, \delta z ^j)$.


\section{Simultaneous Constraints on Cosmology and Photometric Redshift Bias}\label{section:simultaneous_constraints}

Fig. 3a shows constraints on the matter density $\Omega_{\rm m}$ and clustering amplitude $\sigma_8$ for various assumptions about the weak lensing photometric redshift uncertainties. 
The green dashed lines are from the fiducial analysis assuming the weak lensing redshift distributions are known precisely ($\Delta \delta z = 0$).

We investigate the effect of using an incorrect redshift distribution in the simulation, expanding on a similar technique developed in \citet{bonnettsv15}.
We bias the $n(z)$ in the simulated data vector by 0.05 in redshift, a value inspired by the bias calibration applied to the \blockfont{bpz} redshift distribution in \citet{bonnettsv15} to match simulations. 
We erroneously continue to assume the redshift distributions are perfectly known. The result (blue dotted line, Fig. 3a) is now incompatible with the true cosmology at greater than 95\% confidence. Finally we allow freedom in the value of the \pz biases $\delta z^i$. Specifically we marginalise over the fiducial prior of width $\Delta \delta z =0.1$. The magenta solid contours are shifted back close to the true input cosmology, despite the erroneous redshift distributions being used. The width of the magenta contours is not greatly degraded relative to the case where the redshift distributions are perfectly known (green dashed and blue dotted). Quantitatively we find the error on $S_8$ is degraded by 40\%. For Fig. 3b we carry out the same calculation as in Fig. 3a except we additionally vary the dark energy equation of state $w_0$; qualitatively similar results are obtained.

In Fig. 4 we investigate in more detail how the prior width $\Delta \delta z$ on the redshift distribution bias $\delta z^i$ affects the uncertainty $\sigma S_8$ on $S_8$. We contrast the results from cosmic shear alone (blue) with those from the combination of cosmic shear, galaxy clustering and shear-density cross-correlations (magenta). We show results using our fiducial systematics assumptions (magenta) and using less conservative assumptions (green). Specifically we assume no multiplicative shear calibration uncertainty and no intrinsic alignment uncertainty. We see clearly that cosmic shear alone cannot self-calibrate photometric redshift uncertainties, whereas the combination with galaxy clustering and shear-density cross-correlations weakens the dependence of the constraint on prior width, for both fiducial (magenta) and optimistic (green) systematics.
$S_8$ is significantly biased ($\delta S_8$) at all values of $\Delta \delta z$ for cosmic shear (dotted lines), and is biased very little by $\Delta \delta z=0.1$ for the combination of datasets (dot-dashed).

Fig. 5 gives some insight into how the self-calibration works and uses the erroneous biased input redshift distribution (\blockfont{skynet}$- 0.05$) as an illustration. The blue (dot-dash) contours show the degeneracy between cosmology and photometric redshift uncertainties from cosmic shear alone. The contours are only closed because we have applied a conservative prior on photometric redshift uncertainties ($\Delta \delta z=0.1$). In the absence of additional information the cosmology constraints from cosmic shear will be biased because the prior on $\delta z^i$ is centered on zero whereas the truth is at $\delta z^i = 0.05$. The galaxy clustering and shear-density cross-correlations constrain a different degenerate combination of cosmology and redshift bias (pink dotted). Thus when these three two-point functions are combined they produce the purple (solid) contours, which are now centered close to the true cosmology and have a much smaller uncertainty on cosmology or photometric redshift uncertainties than either cosmic shear alone or galaxy clustering plus shear-density cross-correlations alone.

 \begin{figure}
   \centering   
   \includegraphics[width=\columnwidth]{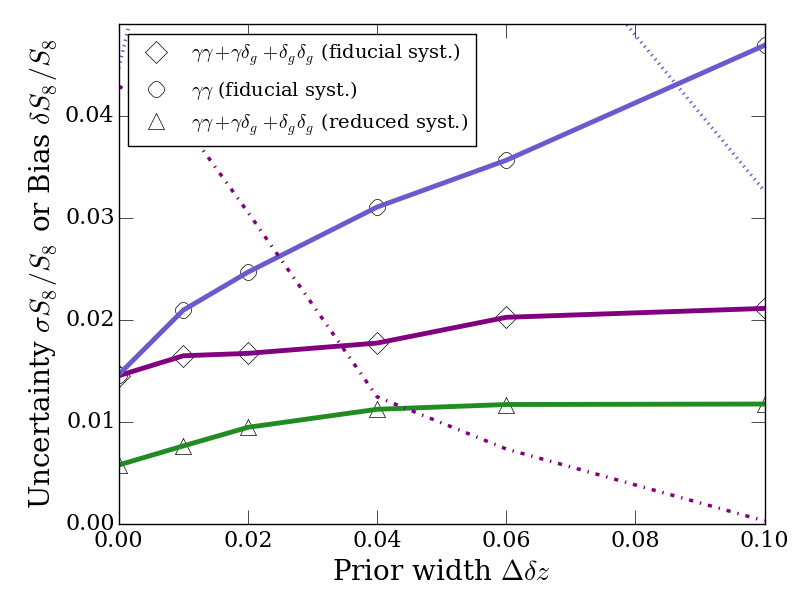} 
\caption{
Uncertainty $\sigma S_8$ on $S_8 \equiv \sigma_8 (\Omega_{\rm m}/0.31)^{0.5}$ 
for different prior widths $\Delta \delta z$ on the redshift distribution biases $ \delta z^i$. 
Blue circles (joined by solid lines) show cosmic shear alone and diamonds (purple solid) denote cosmic shear combined with galaxy clustering and shear-density cross-correlations. The triangles (green solid) are the latter combination, but
assuming intrinsic alignments and multiplicative shear calibration are perfectly known. In the first two cases we also show the bias $\delta S_8 $ induced by marginalising with an (erroneously zero centred) prior of width $\Delta \delta z$ as dot-dashed and dotted lines.
}
\label{fig:ds8s}
\end{figure}

 \begin{figure}
\includegraphics[width=1\columnwidth]{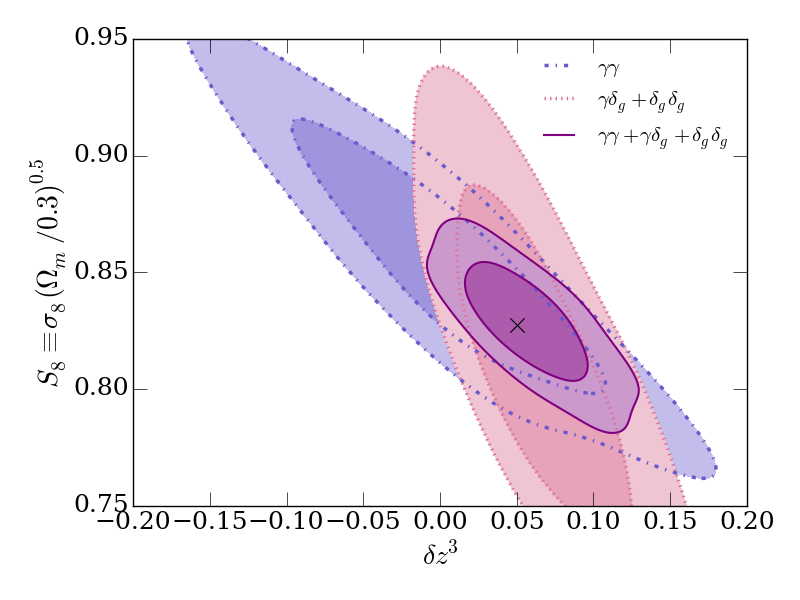} 
\caption{ 
Degeneracies between \pz bias in the furthest redshift bin ($\delta z^3$) and cosmology ($S_8$) 
for cosmic shear alone (blue dash-dot), galaxy clustering and shear-density cross-correlations (pink dotted) and the combination (purple solid).
The input parameters are shown by the black cross. }
\label{fig:nofzs}
\end{figure}

\begin{figure*}
   \centering
   (a)
\includegraphics[width=1\columnwidth]{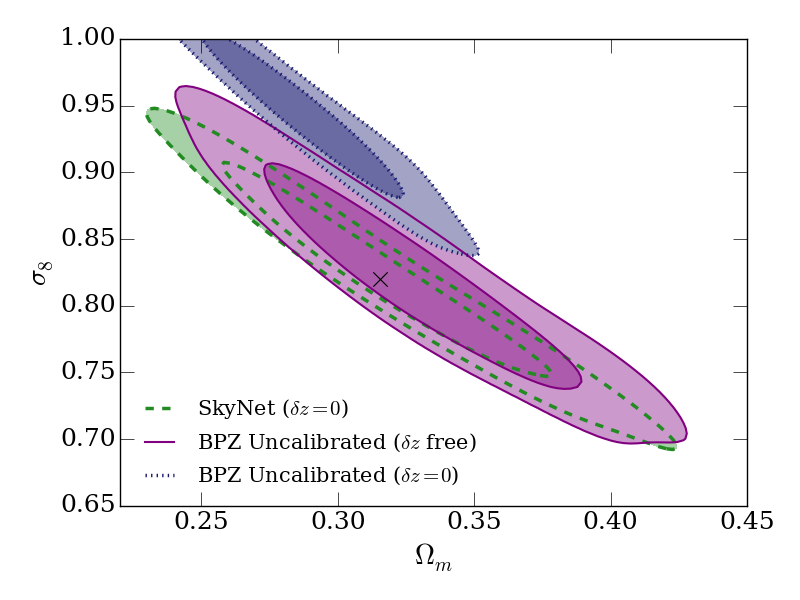}
(b)
\includegraphics[width=1\columnwidth]{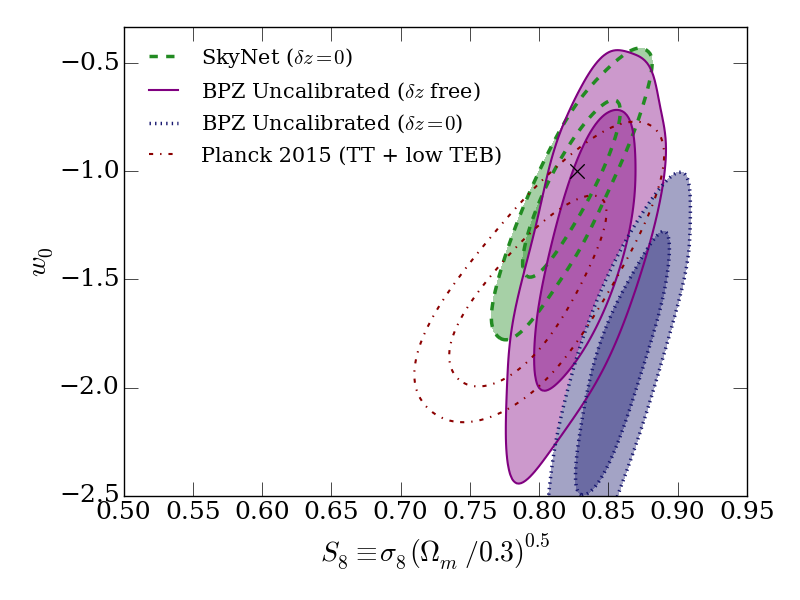}
   \caption{As Fig. \ref{fig:biases} but now using a more realistic realisation of the discrepancy between the estimated and true galaxy redshift distributions. Here we use the $n^i(z)$s from an alternative photo-$z$ code (\blockfont{bpz}) in the simulated data vector and the $n^i(z)$s of the fiducial photo-$z$ code (\blockfont{SkyNet} in the fit.
}   
\label{fig:bpz_constraints}
\end{figure*}

We investigate the robustness of these results to perturbing the fiducial assumptions and calculate the degradation 
$D \equiv \sigma S_8 (\Delta \delta z = 0.1)/\sigma S_8 (\Delta \delta z = 0) -1= 40\%$ for the fiducial analysis.
We find that introducing stochastic bias with one free scale-independent parameter per redshift bin $|r^i_g|<6$ makes little difference ($D=45\%$), but increases $\sigma S_8(0)$ by 6\% relative to the fiducial case. Marginalising over an additional photometric redshift uncertainty parameter per bin, which stretches the redshift distributions $\tilde{n}^{i}(z)=n^{i}(z+S^{i}_z [z-z_p])$, increases the uncertainty on cosmology for all values of $\Delta \delta z$ and thus reduces the relative degradation ($D=28\%$). 
A similar reduction ($D=24\%$) occurs if 
we increase $\ell_{max}$ by a factor of 3.
We also rerun our analysis using a wide prior on the shear measurement bias ($\Delta m=0.05$), and find a degradation $D=16\%$, 
due to a $\sim72\%$ degradation in the error on $S_8$ independent of $\Delta \delta z$.
The biggest impact arises from using a more conservative value for the prior on the photometric redshift uncertainties of the galaxy clustering sample ($\Delta \delta z =0.05$), which breaks the self-calibration, giving a factor of two degradation ($D=103\%$), with $\sigma S_8(0.1)$ increased by 75\%, and $\sigma S_8(0)$ increased by significantly less (6\%), relative to fiducial. In all instances considered, we find no significant residual bias $\delta S_8$ when  $\delta z^i$ are marginalised.

Finally we test the self-calibration result using a more realistic situation in which the true redshift distribution differs from the assumed distribution by more than a simple uniform bias. To illustrate this situation we take the $n(z)$ from DES SV from an alternative photometric redshift code (\blockfont{bpz}). Of the SV codes, \blockfont{bpz} was the most discrepant from our fiducial \blockfont{skynet} choice. We additionally have not applied the 0.05 shift derived from simulations and used in the DES SV analysis, to provide a relatively stringent test. 
Fig. 6 shows the result in the same format as Fig. 3 By construction the green contours in Fig. 6 are the same as those in Fig. 3, and use the correct fiducial distribution in both the simulation and the fit. The blue contours use the qualitatively different \blockfont{bpz} $n(z)$ in the simulation and assume the fiducial \blockfont{skynet} ($\delta z^i = 0$) redshift distributions are perfectly known ($\Delta \delta z=0$) in the fit. The results are now biased by more than in Fig. 3. The result of marginalising over uncertainties in the bias in the redshift distributions (prior width $\Delta \delta z=0.1$) does not trivially move the contours (magenta) back onto the input cosmology (black cross) this time. This suggests that the simple uniform bias in redshift might not always sufficiently account for the differencences between the true and estimated distributions, depending on the survey specifications. In this particular case the truth is still within the $68\%$ confidence contour, but in general a more detailed investigation would be necessary taking into account the specifics of the possible range of redshift errors. 


\section{Conclusion}

We have investigated the potential for current cosmological galaxy imaging surveys to self-calibrate photometric redshift distribution uncertainties, for the first time considering the realistic case in which the weak lensing sample is different from the galaxy clustering sample, 
and has substantial calibration uncertainties. 
We focus on a preliminary Stage III dataset with $\sim 40$M galaxies, in which the galaxy clustering sample has well-understood photometric redshifts ($\Delta \delta z=0.01$).

We find that the combination of cosmic shear, galaxy clustering and shear-density cross-correlations is much more robust to errors and uncertainties in the redshift distribution calibration than cosmic shear alone. Specifically, the uncertainty on the clustering amplitude parameter $S_8 \equiv \sigma_8 (\Omega_{\rm m}/0.31)^{0.5}$ is increased by only $40\%$ on marginalising over three free independent bias parameters with a prior width $\Delta \delta z=0.1$, relative to the case $\Delta \delta z=0$. This contrasts with more than a factor of two degradation for cosmic shear alone. We illustrate that this is because cosmic shear constrains a different degenerate combination of cosmology and photometric redshift calibration parameters than galaxy clustering and shear-density cross-correlations. 

We find that the combination of all three two-point functions can correct even a substantial bias (of 0.05) in the $n(z)$ to accurately recover the input cosmology. This result is robust to allowing for a basic stochastic bias model, and strengthened by using less conservative cuts on the scales used in the galaxy clustering analysis. The self-calibration result disappears if the redshift distributions are less well understood for the galaxy clustering sample ($\Delta \delta z=0.05$). Using an alternative redshift distribution estimate  (\blockfont{bpz}) we demonstrate that this result may change if the deviation of the redshift distribution from the truth is not fully captured by a uniform translation, if the only redshift uncertainty considered is a uniform translation in redshift. The validity of our findings should be verified on a case-by-case basis for specific realisations of the photo-z error

This investigation advances on most previous numerical forecasts in implementing MCMC sampling rather than Fisher analyses, and assumes a low-density galaxy clustering sample with relatively well-known redshifts. We do, however, assume Gaussian covariance matrices,
which tend to underestimate the uncertainties for cosmic shear and could thus make our forecasts over-optimistic. Investigation of non-Gaussian covariances is beyond the scope of this Letter. We have also assumed the Limber approximation is correct for the range of scales used, and ignored redshift-space distortions.
The results suggest that self-calibration may be a practical solution for current cosmological surveys, assuming reliable \pz estimates can be obtained for the galaxy clustering catalogue, if the weak lensing redshift distributions cannot be easily calibrated through a different route.

\section*{Acknowledgements}
We thank Scott Dodelson, Enrique Gaztanaga, Ashley Ross, Martin Crocce, Eduardo Rozo, Jonathan Blazek, Adam Amara, Benjamin Joachimi, Catherine Heymans, Rachel Mandelbaum, Josh Frieman and the DES Multi-Probe Pipeline group for helpful conversations.
SS recognises receipt of a UK Science and Technology Facilities Council (STFC) Doctoral Training Grant. SLB, MT, JZ and DK acknowledge support from the European Research Council.

\bibliographystyle{arxiv}
\bibliography{samuroff.bib}
\label{lastpage}

\end{document}